# C-RED 3: A SWIR camera for FSO applications

J.L. Gach*[a,b], David Boutolleau [a], Cecile Brun [a], Thomas Carmignani [a], Fabien Clop [a], Philippe Feautrier [a,c], Stephane Lemarchand [a], Eric Stadler [a,c], Yann Wanwanscappel [a]

[a] First light Imaging S.A.S., Europarc Ste Victoire, Route de Valbrillant, 13590 Meyreuil, France;
[b] Aix Marseille Univ, CNRS, CNES, LAM, Marseille, France;
[c] Univ. Grenoble Alpes, CNRS, IPAG, F-38000 Grenoble, France;

## ABSTRACT

Free space communications (FSO) is interesting for distant applications where high bandwidth is needed while using a fiber is not possible. However these links have to face several issues, and the most important one is the beam scintillation due to the propagation through a turbulent media, the atmosphere. Several mitigation strategies have been developed, but the best way to suppress scintillation is to use adaptive optics, widely used now in astronomy. The main difficulty for FSO is to probe the wavefront fast enough to have a good turbulence correction. This was not possible due to the lack of wavefront sensors working in the SWIR. C-RED 3 is a 640x512 SWIR camera running at 600FPS full frame and has the legacy of all the developments of astronomical infrared fast wavefront sensors on top of specific features for FSO (Low SWaP, Low Cost). We will present the performances of this new camera and demonstrate how it fulfills the needs of FSO adaptive optics.

## 1. INTRODUCTION

Free space communications are good candidates to improve the throughput of the communication systems when using a fiber link is not possible. This is the case for airborne systems which are limited now by the radio links, since it is commonly admitted that the maximal throughput is in the order of 10% of the baseband frequency. Using the Ka band (20 to 40 GHz) offers enhanced transmission throughput, but still limited compared to what is possible with an optical communication. And this is becoming an increasing problem with the increasing amount of data to exchange in particular with satellites. FSO communications are also interesting when security is important since the optical beam can be very directive and impossible to "listen" without interfering and then indicating that the transmission is compromised.

Light propagating through the atmosphere is known to be disturbed by atmospheric turbulence. This is a well-known phenomenon of astronomers, and because of this the telescopes are limited in spatial resolution to the "seeing" which is correlated to the observing site and not the instrument diffraction limit which is directly linked to the main pupil size. In the best astronomical sites, telescopes of 40cm size have the same spatial resolution as several meter telescopes.

To overcome this the idea of correcting in real time the atmospheric turbulence was firstly introduced by Babbock as early as 1953 [1], but at that time the technology was clearly not there. Later in 1977 Hardy at al. worked on this concept for military applications [2], but the real large scale development of this technique occurred in astronomy with the work of Rousset at al. at Haute Provence Observatory in 1990 [3] and Rigaut et al. at the 3.6m ESO telescope in 1991 [4] both with the COME-ON experiment. Since these works, numerous groups have developed adaptive optics systems for various telescopes and instruments. Adaptive optics systems started to become very popular and quite easy to make, with the progresses of wavefront sensors, deformable mirrors and real time computers which are the key elements of an adaptive optics loop. Now it is used also in ophthalmology for high resolution retina imaging, and also in microscopy to enhance the spatial resolution.

For FSO applications, the main issues introduced by turbulence is the ability to inject the beam energy into a single mode fiber, to go to the detector. But the turbulence will create mainly scintillation due to the beam speckles, and for small aperture telescopes, it may also create some beam position jitter due to the residual wavefront tip-tilt.

*jeanluc.gach@first-light.fr, www.first-light-imaging.com

## 2. THE C-RED 3 CAMERA

**2.1 Astronomical needs for wavefront sensing**

The key parameter of an adaptive optics system is the wavefront sensor and its ability to give an instantaneous picture of the incoming wavefront. It is usually made of a camera and an optical system to probe the wavefront on several points of the pupil. The most popular ones are the Shack-Hartmann sensor which consists of a lenslet array placed in a conjugate plane of the pupil or the pyramid sensor which is in the image plane or one of its conjugates.

Since the beginning of adaptive optics systems, these wavefront sensors were mainly used in the visible. This comes from the camera that was available in the visible whereas in the infrared no suitable camera was available. Indeed for this application, several 100s of Hz of framerate are necessary with a good readout noise at the same time. Our team worked a lot on the improvements of visible cameras for wavefront sensing with the OCAM² which is to date the fastest and lowest noise camera, tailored for this application [5][6].

But astronomers were quite frustrated not being able to use the infrared spectrum for adaptive optics applications. Since they are mostly using natural stars, the sky coverage is much more important in the infrared and stars are brighter, so there was a real interest to go to longer wavelengths. Our team decided then to investigate the possibility of building the infrared counterpart of OCAM², which ended with the C-RED One camera using for the first time in a commercial product the electron initiated avalanche photo diode arrays (e-APD) technology [7]. For the first time it was possible to observe in the infrared with KHz frame rates (3.5 kHz) and subelectron (0.5 e-) readout noise at the same time [8]. This was an outstanding progress for AO applications and opened the possibility for wavefront sensing in the infrared. However, this camera while outperforming all others in terms of speed and noise, is a bit bulky and is suitable where the ultimate performance is needed. So we've decided to port the legacy of C-RED One to the InGaAs technology, this gave the C-RED 2 camera which has a 600 FPS frame rate and 30e- readout noise, but still cooled for astronomical applications where intermediate performance is needed (low order wavefront sensing, tip-tilt sensing)[9]. The cooling was necessary because when the guide star is faint, astronomers just slow down the camera to integrate more photons, therefore they use the camera in a speed range where dark current contribution can become significant. This is not the case for FSO application where the downlink power is constant, therefore the cooling is no more needed. This leads to the C-RED 3 camera where all the cooling system has been removed and the electronics squeezed to give a very small high-speed SWIR camera optimized for wavefront sensing at high speed.

For FSO application, the availability of low noise fast infrared cameras to build wavefront sensors changes somewhat the paradigm. It is then possible to use the main beam which is likely at 1.55 µm with a minimal energy pickup, leaving the most part of the energy to the link. Moreover, the low size, weight and power (SWaP) opens the possibility to use this technology on airborne material (planes, UAVs).

**2.2 Camera presentation**

C-RED 3 is a high performance, high speed low noise camera designed for Short Wave InfraRed imaging based on the tecless version of the SNAKE detector from Lynred (formerly Sofradir) [10], [11], [12], [13], [14], [15]. C-RED 3 integrates a 640 x 512 InGaAs PIN Photodiode detector with 15 µm pixel pitch for high resolution, which embeds an electronic shutter with integration pulses shorter than 10 µs. C-RED 3 is also capable of windowing and region of interest (ROI), allowing faster image rate while maintaining a very low noise which is very interesting for FSO applications where a limited number of modes will be corrected, so a smaller amount of pixels needed. At the same time the frame rate increases in ROI mode up to several kHz (for example 4.8 kHz in 128x128 and 9.5 kHz in 64x64 ROI respectively) which is also interesting for FSO with LEO satellites where the apparent wind is high due to the satellite speed and therefore a high framerate correction needed.

The software allows real time applications, and the interface is CameraLink full and superspeed USB3. C-RED 3 is designed to be updated remotely. The camera can operate in very low-light conditions as well as remote locations. Designed for high-end SWIR applications, smart and compact (see Figure 1), C-RED 3 is operating from 0.9 to 1.7 µm with a very good Quantum Efficiency over 70%, offering new opportunities for industrial or scientific applications.

The camera exists also in OEM version without housing and with custom features if needed. It has been designed for low SWaP with 55x55x60 mm³ dimensions, 230g in standard version and 100g in OEM version. Figure 1 shows the camera and Table 1 its main features.

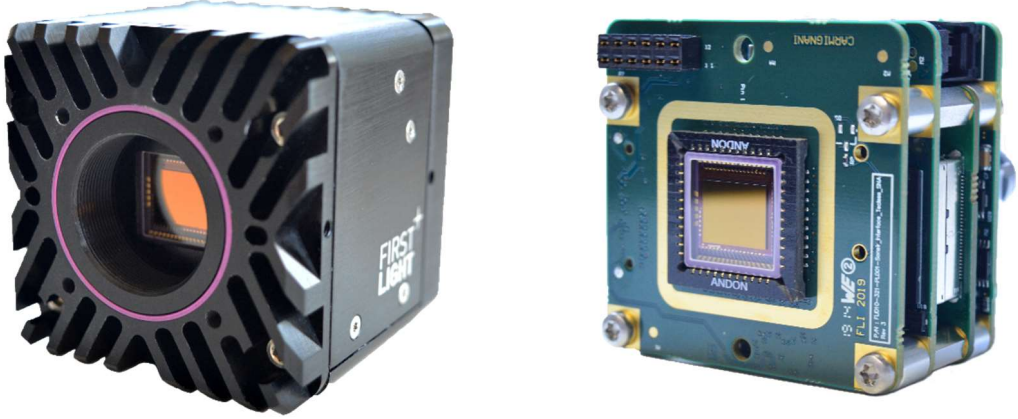

*Figure 1: Picture of a C-RED3 camera (left) and OEM module (right). The actual size is 55x55x60 mm*

The Table 1 summarizes the main features and performances of the C-RED 3 camera.

| Test measurement | Result | Unit |
|---|---|---|
| Maximum speed | 602 | fps |
| Mean dark + readout noise at 600 fps | < 50 | e⁻ |
| Quantization | 14 | bit |
| Quantum efficiency from 0.9 to 1.7 µm | > 70 | % |
| Operability | > 99.7 | % |
| Image full well capacity at low gain, 600 fps | 1400 | ke⁻ |
| Image full well capacity at high gain, 600 fps | 33 | ke⁻ |
| Maximum speed 32x4 window | 32066 | fps |
| Maximum speed 320x256 window | 1779 | fps |

Table 1: typical performances and main features of the C-RED 3 640x512 InGaAs SWIR camera.

## 2.3 Quantum efficiency

The InGaAs diodes are grown with a MOVPE process on an InP substrate which is not removed. Therefore, this sensor has a cutoff wavelength driven by the InGaAs bandgap and a cut-on wavelength driven by the InP bandgap. These cut on and cutoff wavelengths are slightly moving with temperature variations and the popular Varshni model [16] can be applied to compute these variations using the following equation :

$$E_g(T) = E_g(0) - \frac{\alpha T^2}{T + \beta}$$

Various authors give the constants for InGaAs for example Gaskill who measured $E_g(0)$=803 meV, α=4.0×10$^{-4}$ eV K$^{-1}$, and β=226 K [17] and $E_g(0)$=1.424 eV, α=3.91×10$^{-4}$ eV K$^{-1}$, and β=243 K for InP reported by Passler [18].

Moreover, the InP substrate becomes more transparent with decreasing temperature, this is due mainly to the refractive index variation and absorption coefficient variation with temperature [19]. The total effect of temperature is then a slight shift of the wavelength responsivity towards the red end of the spectrum and a slight QE decrease when the temperature is increasing. However, these effects can be considered as negligible for a normal utilization, and Figure 2 shows the measured QE curves vs temperature variation.

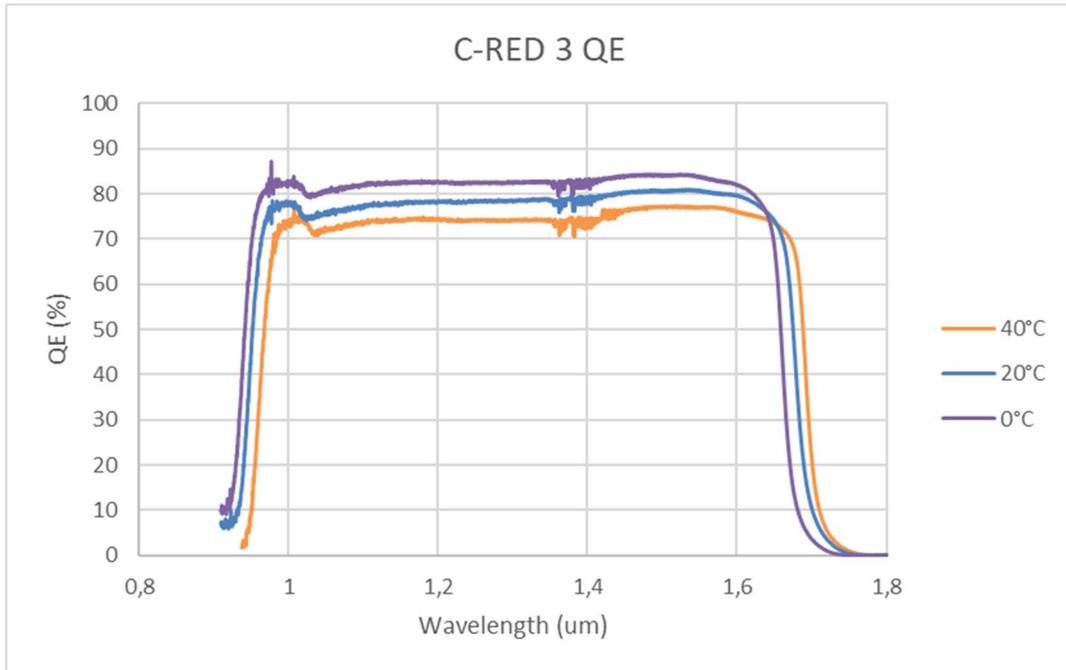

*Figure 2: C-RED 3 QE response vs wavelength and operating temperature*

## 2.4 Dark current

Since the camera is uncooled, the dark current is rather high. But at the same time the frame rate is also high, therefore the elementary image dark contribution is maintained to a low level. Dark current contributes to two unwanted effects in the image. Firstly, it is a background signal that fills up the pixel capacity, therefore it limits the useable well capacity. Secondly it is also a source of noise, because on each image, the shot noise contribution of the dark is in $\sqrt{N}$ where N is the amount of dark electrons.

Figure 3 shows the typical dark current measured on a camera for various temperatures. It follows the Arrhenius law as expected with an exponential increase with temperature. For an ambient temp utilization with 35°C sensor temperature, this gives an impressive 600ke$^-$/s/pixel. However, when scaled to the exposure time at 600FPS this drops to 1000e$^-$/pixel, which is 1/40$^{th}$ of the well capacity in high gain mode and contributes to 30e$^-$ for the readout noise. When used in windowing (ROI) at higher frame rates, this contribution drops to negligible levels (127 e$^-$ at 4.7 kHz for example) validating the concept of an uncooled camera for FSO applications.

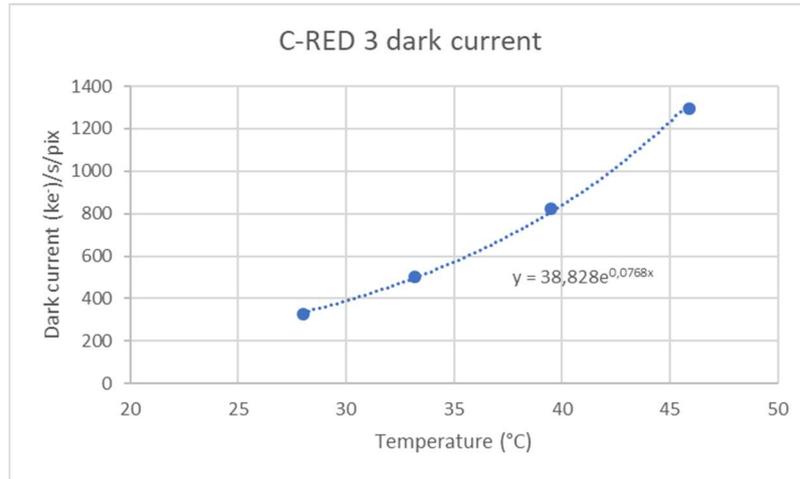

*Figure 3: typical C-RED 3 dark current vs temperature and exponential fit*

## 2.5 Readout noise

The best readout noise is obtained in high gain configuration where the pixel capacity is the lower (~35 ke$^-$). It is as low as 25-30e$^-$ at low integration time (50μs) and increases for longer ones due to the dark current shot noise contribution.

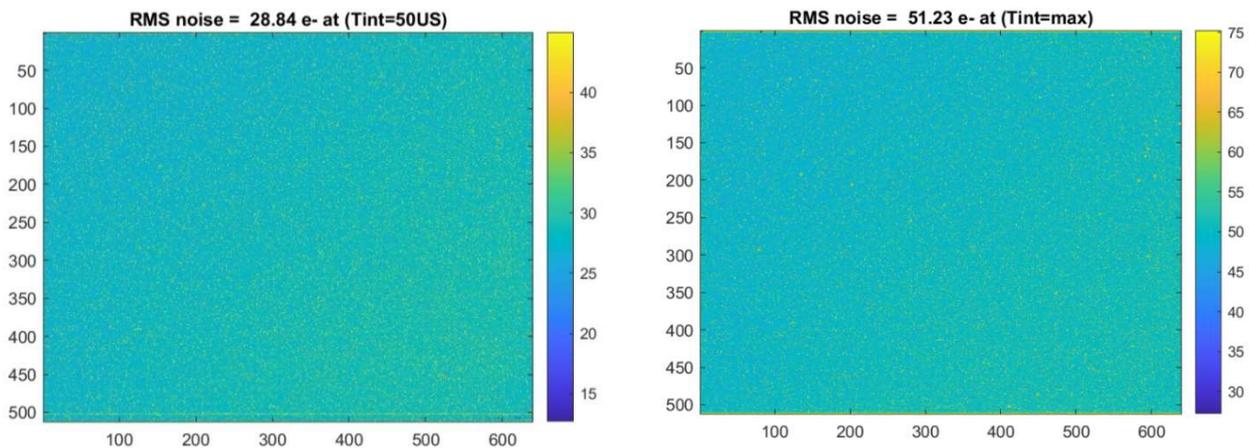

*Figure 4: Typical noise map of a C-RED 3 camera at 50us integration time (left) and 1.6ms (right) with the high gain mode.*

Figure 4 shows a typical noise map of a camera with the high gain mode. At higher integration time, it raises up to 50 e$^-$ due to the dark shot noise contribution that adds in a quadratic way to the readout noise.

When used at mid gain (~110 ke$^-$/pixel), the typical readout noise is 60e$^-$ at 1.6ms integration time, and at the lowest gain setting (1.3Me$^-$/pixel capacity) the noise is typically 300e$^-$. In this mode the dark shot noise contribution is negligible up to integration times of 100 to 200ms. Figure 5 summarizes the operating regimes of the camera vs integration time showing that the uncooled operation is a very valid approach for frame rates faster than 100 FPS in high and med gains and 5 fps in low gain, which is the case for the foreseen application.

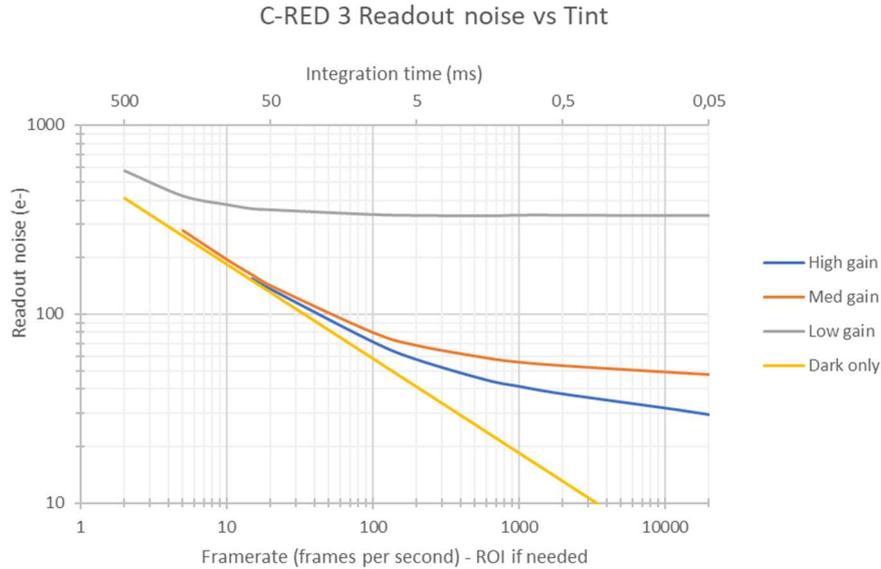

*Figure 5: C-RED 3 readout noise vs integration time at 35°C showing that for speeds above 100 fps in high gain and med gain and 5fps in low gain the camera is readout noise limited. The dark current contribution is negligible. For lower frame rates the camera is dark current noise limited therefore a cooled camera like C-RED 2 is more suitable for the application.*

## 2.6 Linearity

The linearity is an important parameter for wavefront sensing. Usually it must be better than 5%, and preferably below 2.5%. The linearity is measured for each camera and is usually better than 1% for the high gain mode, and 0.5% for the medium gain and low gain. Figure 6 shows the typical linearity deviations.

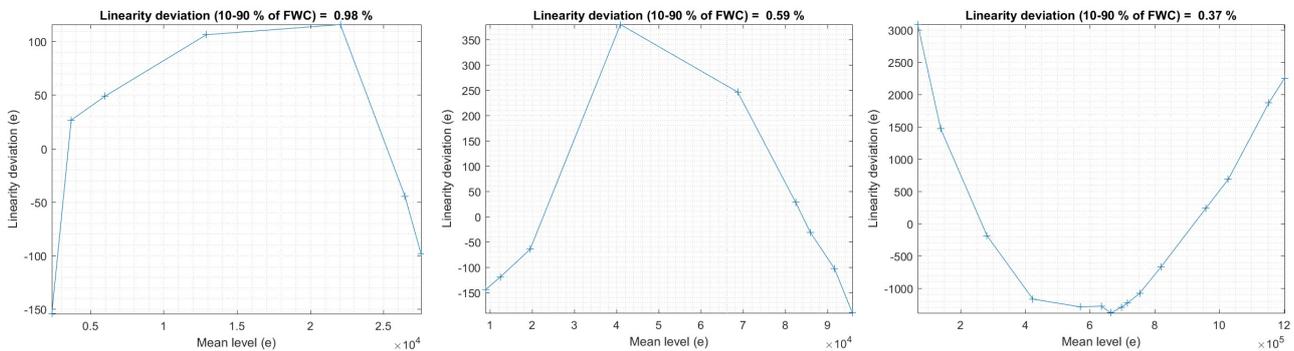

*Figure 6: Typical linearity errors in high gain (left), med gain (center) and low gain (right)*

## 2.7 Camera features

The camera includes a lot of embedded features and on-the-fly correction capabilities by the FPGA fabric and the on-board processor. On top of standard corrections such as bad pixel removal or automatic gain control (AGC), the most important one is the ability to have an adaptative bias+dark correction which is automatically updated according to the operating conditions (exposure time and sensor temperature) because the camera is not stabilized in temperature, therefore the dark rate varies. If the actual conditions varies too much from the previous conditions, then a new correction map is computed by the camera firmware and replaced in the real time subtraction system (see Figure 7), doing so the user has permanently a useable image. This adaptative correction might be disabled and replaced by standard calibrations.

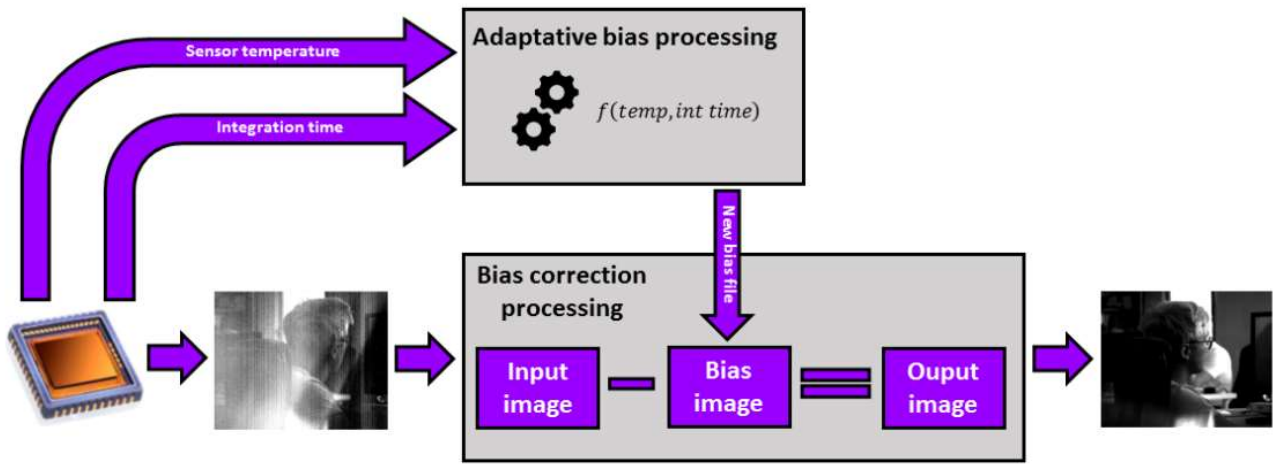

*Figure 7: the adaptative bias correction embedded in the camera*

The camera has also numerous synchronization mechanisms (in and out) so that it can be synchronized to an external process

## 3. CONCLUSION

We bring to the FSO community a new camera originally designed for high performance astronomical adaptive optics applications and adapted to the FSO applications. We made the wager of an uncooled approach for this specific application because it has numerous advantages over cooled or temperature stabilized cameras. With its outstanding performances in terms of speed, readout noise and its low SWaP, we expect that it will become a tool for simplifying adaptive optics applied to FSO links and open new applications for planes and UAVs.